\documentstyle[aps,prl,twocolumn,psfig]{revtex}
\def\bfm#1{\mbox{\boldmath$#1$}}
\begin{document}

\draft

\preprint{Not yet submitted}

\wideabs{
\title{Spectral Properties of Atoms in Fields: A Semiclassical Analysis}

\author{P. N.\ Walker and T. S. Monteiro}

\address{Department of Physics and Astronomy, University College London,
Gower Street, London WC1E 6BT, U.K.}

\date{\today}
\maketitle

\begin{abstract}
 
We develop a semiclassical theory for the spectral rigidity
of  non-hydrogenic Rydberg atoms in electric fields and
evaluate the significant deviations from the well-known 
Poissonian behaviour in the hydrogenic case.
The resulting formula is shown to be in excellent 
agreement with the exact quantal result.
We also investigate diamagnetic atoms; we
find that, in contrast to the classically
integrable atoms, diffraction has a  small effect on the
spectral rigidity in the classically chaotic atom. We show
 our predictions can also be of use in the mixed
phase space regime. 
\end{abstract}

}

The short-range spectral properties of non-hydrogenic
Rydberg atoms in external
fields were recently found to have an unexpected  character ~\cite{Jonck98},
the nearest neighbour distribution was neither Poisson nor Wigner-Dyson,
but close to a new generic {\it intermediate} class known as
{\it half-Poisson}. These findings have been
investigated experimentally for helium atoms in electric fields ~\cite{Kips},
 following  much interest in the dynamics of
non-hydrogenic atoms in weak fields and the effects of {\it core-induced chaos}
~\cite{GD92,DD94,Court94,Hupp95,Dan98}.
Concurrently, there has also been much interest in such
intermediate Nearest Neighbour distributions ~\cite{Bog98,Pasc98} with
broad application in problems such as electron-electron interactions
in closed mesoscopic devices \cite{Pasc98,Yuval},
the metal-insulator transition \cite{Braun}, intruder states in nuclear
physics \cite{Bohr}, and inclusions in billiards \cite{Bog98}.
These studies of intermediate statistics have
exclusively employed quantum calculations.

Here we obtain spectral rigidities of non-hydrogenic Rydberg atoms in electric
and magnetic fields from an accurate quantal calculation.  We find interesting
and substantial deviations from Poissonian (hydrogenic ) behaviour in the
integrable (electric field) and near integrable (weak magnetic field) case.  We
develop a semiclassical theory for Stark spectra which is in good
agreement with the quantal results and shows that one-scatter diffractive
orbits account for most of the effect.  In contrast we find a comparatively small
effect in the case of fully chaotic Rydberg atoms.  To our knowledge this is
the first analysis of the curve-form of the spectral rigidity for a generic
atom, which we derive from classical dynamical information. Our semiclassical
analysis should be extendable to the mixed phase space Kolmogorov-Arnold-Moser
(KAM) case.

The spectral rigidity, defined as
\begin{equation}
\Delta(L)=\min_{A,B}{\rho\over L}
\int_{-L/2\rho}^{L/2\rho}\!\!\!\!\!de
\Big\langle [{\cal N}(E+e)-A-Be ]^2
\Big\rangle_E \ ,
\label{rig}
\end{equation}
where ${\cal N}(E)$ is the spectral staircase function, $\rho$
is the density of states (assumed constant over the range of
$L$ to be considered) and $\langle\dots\rangle_E$ denotes
averaging over the spectrum, was first analysed
semiclassically for classically integrable and chaotic systems
by Berry \cite{Berry85}. It provides a measure of long-ranged
deviations from the Weyl rule for spectral density, and for
diffraction-free systems is controlled (for $L\agt {\cal O}(1)$)
by classical dynamical correlations. For $L\alt {\cal O}(1)$ it converges
universally to the form $\Delta(L)=L/15$ for symmetry reduced spectra. 

The hydrogen atom in a static electric field (of strength $F$)
or in a magnetic field (of strength $B$)  provided some of the {\it cleanest}
illustrations of integrability, mixed phase space and chaos
in a real system ~\cite{Fried} . The  dynamics of the electron is two-dimensional
and has a 
useful scaling property: the classical dynamics depends only on a
scaled energy $\epsilon=E \kappa^2$, where 
$\kappa=F^{-1/4}$ for the electric field case and $\kappa=B^{-1/3}$
for the magnetic field case. This property is exploited in both
experiment and theory. Spectra are obtained at fixed $\epsilon$ and
the corresponding eigenvalues $\kappa_i$ represent effective values
of $\hbar^{-1}$ with fixed classical dynamics. 
Hydrogen in a magnetic field is near-integrable for $\epsilon <-0.5$;
as the field is increased it makes a gradual transition to full
chaos at  $\epsilon \simeq-0.1$. The Hamiltonian
for hydrogen in an electric field is
always separable.  We consider  
field values where the eigenvalues are well below the ionization threshold at $\epsilon=-2$, so the system may be considered integrable and bound.

Most experiments in fact investigate atoms other than hydrogen
(typically He, Li or Rb). The useful scaling
property may still be exploited but the inner core
of electrons yields non-trivial effects: additional weak spectral modulations
and spectral statistics near the Wigner-Dyson limit
even for the integrable/near integrable regime 
~\cite{DD94,Court94,Hupp95}. Hence the interest in so-called
{\it core-induced chaos}. 
The additional modulations are accurately described by Diffractive 
Periodic Orbit Theory ~\cite{Dan98}. An investigation ~\cite{Jonck98} of the
NNS statistics for the lowest 40,000 eigenvalues
showed that  they are only near Wigner-Dyson for the
lowest $\sim 1000$ states. For small $\hbar$ they were found
 to make a transition to an intermediate distribution 
 $P(s) \sim \alpha s e^{-2s}$ with ($\alpha \sim 3-4$),
  near to the Half-Poisson distribution (for which $\alpha =4$).

Rydberg atoms and molecules are well described by
Quantum Defect Theory, one of the most widely used theories
in atomic physics \cite{Fried,GD92}. The core is described by a set of phaseshifts
(quantum defects) $\delta_{l}$ in each partial wave,
quantifying the departure from pure Coulomb behaviour.
We consider the s-wave scattering
case, where $\delta_0$ is the only non-zero phaseshift,
which describes lithium ($\delta_{0} \simeq 0.41\pi$) and helium
($\delta_{0} \simeq 0.3\pi$ for triplet and 
$\delta_{0} \simeq 0.14\pi$ for singlet helium ) extremely well.

The insertion of a single scattering
channel can be represented as a perturbation by a projection operator.
In this case, it has been shown \cite{Jonck98,Bog98,Pasc98,Bohr} that
the eigenvalues of the perturbed system remain trapped between the
unperturbed eigenvalues. This trapping puts a strong restriction
on perturbations to long-range spectral correlations; indeed, in
the limit $\hbar\to 0\ ,L\to\infty$, the perturbed spectral rigidity
can differ from the unperturbed one by at most some value, bounded
by 0 and 2.

\begin{figure}[htpb]
\centerline{
\psfig{figure=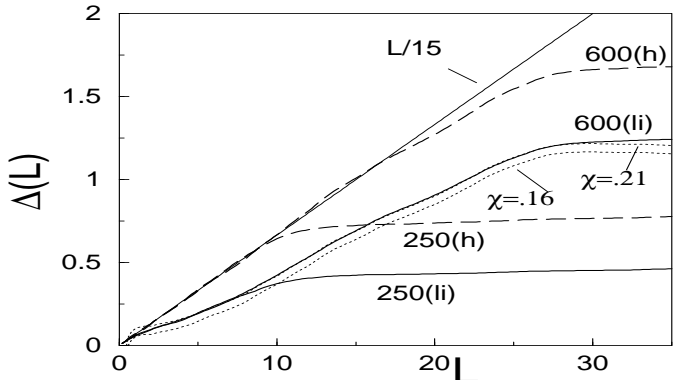,height=2.in,width=3.5in,angle=270}
}
\caption { Rigidities calculated from quantal spectra of
 hydrogen  and  lithium-like atom
for $\epsilon=-3$.
The effective values of $\hbar^{-1}$ (ie the $\kappa$)
 are indicated. The straight line indicates the $L/15$ Poisson 
limit.  The figure shows that the non-hydrogenic rigidities follow a 
significantly different 
curve. Surprisingly, given the $\hbar$ dependent nature
of the diffractive process, the perturbed curves are essentially
  independent of $\hbar$ below the saturation 
value of $L \sim L_{\rm max}$. The curves from our final
 semiclassical formula,Eq.(9), corresponding to both the SC ($\chi=.16$)
 and QM ($\chi=.21$) values of
 the constant shift are superposed on the $\kappa=600$ non-hydrogenic quantal
rigidity, showing the excellent agreement and 
the relatively modest effect of the uncertainty
in the shift.}
\label{Fig1}
\end{figure}

We have calculated the lowest $36000$ states of the Stark spectrum
for magnetic quantum number $m=0$, $\epsilon=-3$ 
and a range of quantum defects. The results for
$\delta_0=0$ (hydrogen) and $\delta_0=\pi/2$ ('lithium') for
values of the effective $\hbar^{-1}$, $\kappa <600$ are 
shown in Fig ~\ref{Fig1}.
These values span and go well beyond typical experimental values.
( the experimental NNS statistics ~\cite{Kips} correspond
to $\kappa \sim 130-150$). Fig ~\ref{Fig1} shows that for our $\kappa <600$
range, deviations from Poissonian/hydrogenic
 ($\delta_0=0$) behaviour are substantial. Even for such small values of 
 the effective $\hbar$ though, it is evident that the perturbed curves are not
 obtained by a simple constant shift.

 The electronic core yields a combination of  
Coulomb and s-wave scattering. The Coulomb scattering generates
Gutzwiller Periodic Orbits (POs): {\it geometric orbits}.
The effect of the s-wave ($\delta_0$) scattering is to generate
 {\it diffractive} trajectories,  ${\cal O}(\sqrt {\hbar})$
in amplitude weaker than isolated geometric POs. For atomic core-scattering
 orbits which close at the nucleus corresponds to 
either  periodic  or 
half-periodic orbits : hence -for chaotic or regular dynamics-
every diffractive orbit resulting from a
single scattering is paired with a
 geometric periodic orbit or a half-periodic
 orbit of the same action. In the Stark  case there are no half-period
contributions; in the magnetic field  case there are half-period
diffractive contributions (the 'D' orbits seen in ~\cite{Dan98}).
All multiple scattering diffractive orbits
can likewise be associated with a given combination
of geometric periodic or half-periodic orbits. In contrast, in billiard systems
 diffractive orbits are generally unrelated to the geometric periodic orbits.
Their proliferation relative to geometric orbits in the chaotic regime is not 
restricted by this 'pairing'.
 It results in a non-vanishing semiclassical contribution \cite{sieber} .

For  integrable atomic spectra, the geometric contribution to
the staircase function ${\cal N}_G (\kappa) = 
{\cal N}^{\rm s}_G (\kappa) + {\cal N}^{\rm osc}_G (\kappa)$   
includes a sum over contributions \cite{BT} from
integrable tori. In the  scaled atomic spectra,
${\cal N}^{\rm osc}_G (\kappa)  \sim 
\sqrt{\kappa} \sum_j   ( {\cal A}_j / T_j)
e^{i( S_j \kappa  - \alpha_j \pi/2)}\ $
The  amplitudes ${\cal A}_j$ are the Berry-Tabor amplitudes for resonant tori 
which, in action angle variables is
${\cal A}_j=\sqrt{2\pi/|\bfm \omega.\partial \bfm I_j/\partial T_j |
{\rm Det}\{\partial\bfm\theta/\partial \bfm I_j\}}$

All  tori contain just one PO that
collides with the nucleus. Trajectories that can diffract are
therefore isolated in the usual sense. The amplitude $A_D$
of the contribution of a particular single-scatter isolated diffractive
trajectory of action $S$, in our $2D$ system ,was given in \cite{Dan98}
$ A^j_D =  S_j /\pi \sqrt {2 \pi/m_{12} \kappa}
\sin \delta_0 e^{i\delta_0} \sin \phi $
where $m_{12}$ is an element of the reduced monodromy matrix ${\bf M}$,
$\phi$ is the angle of incidence
of the orbit relative to the field direction. 
In integrable systems $m_{12}\propto S$. Hence $A_D$ scales as
$\sqrt{S / \kappa}$. We investigate the ratio of the diffractive
to corresponding geometric contribution, 
$A^j_D/A^j_G= A_D  T_j/( \kappa  {\cal A}_j )$ which we write as
$A^j_D/A^j_G =i C_j \sin \delta_0 e^{i\delta_0}  S_j / \kappa$. By a
unitary transform to action angle coordinates we can show \cite{Walker} that 
$C_j \sim \omega^{(j)}_{1} \omega^{(j)}_{2} $ where the $\omega$
are the frequencies of motion along the two independent
degrees of freedom. 
\begin{figure}[htpb]
\centerline{
\psfig{figure=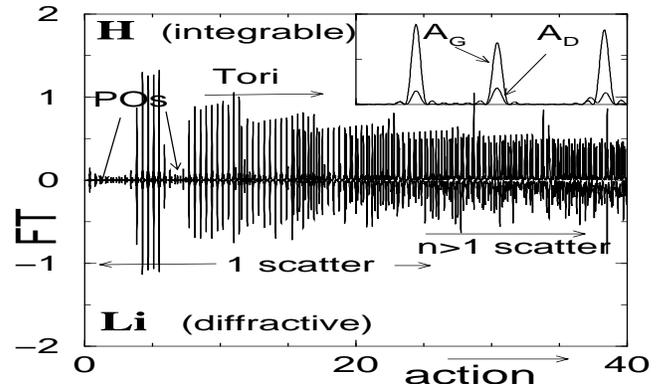,height=2.in,width=3.5in}
}

\caption {
 Fourier transforms of scaled spectra
for hydrogen and a lithium-like atom($\delta=0.5\pi$) for $\epsilon=-3$
from a fully quantal calculation. The inset amplifies the region 
around $S \sim 17$.  
The amplitudes of the diffractive orbits may be obtained from the
(complex) difference of the hydrogen and lithium traces.
The amplitudes of multiple-scattering diffractive orbits
${\cal O}(\hbar^{n/2})$ are relatively weak, but
in the very large $ S$ regime $(L \sim {\cal O}(1))$ the
proliferation in the number of  $n-$ scatter orbits with increasing
action means that the $n=1$ contribution ceases to be predominant.}

\label{Fig2}
\end{figure}

Hence the $C_j$ are independent of
$\kappa$ and do not scale with orbit length. They are nearly constant and
fluctuate weakly about an average value \cite{Walker}.
From the Fourier transforms of scaled spectra , we 
can confirm this behaviour. The $A_D/A_G$ ratios and hence
the $C_j$ statistics are obtained numerically from the first $\sim 50$ pairs.

Since the ratio of the one-scatter diffractive orbits
to the resonant tori contributions is ${\cal O}(S/ \kappa)$,
which in the unscaled spectra is ${\cal O}(\hbar T)$,
on classical timescales diffraction has
a very small effect, but on quantum timescales ($T\sim 1/\hbar$) both
contributions are of the same order. We 
show that a semiclassical analysis of the rigidity is sufficient to
reproduce such essentially quantum phenomena. 

\begin{figure}[htpb]
\centerline{
\psfig{figure=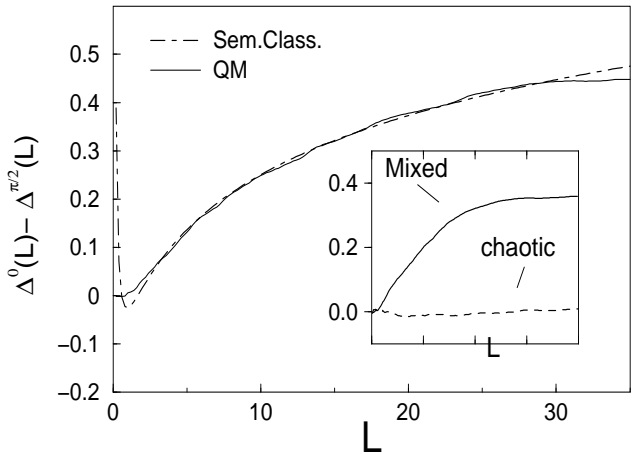,height=2.5in,width=3.25in,angle=270}
}

\caption {Comparison between exact diagonalisation (QM) and
Eq.(\ref{final}) (SC) for the integrable case of atoms in electric fields
for scaled field $\epsilon=-3$ for $\kappa \sim 600$, $\chi=.21$. In order to
expose the approximately logarithmic nature of the diffractive correction
here we plot the {\em difference} between the hydrogenic and diffractive
rigidities. The inset shows the corresponding (quantal only)
results for atoms in magnetic fields for
the near-integrable/mixed case (solid line,
$\epsilon=-.6$,$\kappa \sim 150$) which is qualitatively
similar to the integrable case and the fully
chaotic rigidity differences (dotted line
$\epsilon=-.1$,$\kappa \sim 125$) which are $\sim 0$.}
\label{Fig3}
\end{figure}

We have also calculated the lowest $10000$ states of the diamagnetic problem
$\epsilon=-0.6$ (near-integrable), and
$\epsilon=-0.1$ (chaotic). The results are in Fig ~\ref{Fig3} (inset). 
In the chaotic case we find almost no perturbation
of the hydrogenic result.Isolated hyperbolically unstable
orbits contribute like $|2-{\rm Tr}\,{\bf M}|^{-1/2}$, which vanishes
exponentially with the orbit period. In such a system, $m_{12}$ typically
diverges with the same Lyapunov exponent as ${\rm Tr}\,{\bf M}$. 
Hence $A^j_D/A^j_G$ is ${\cal O} (\sqrt{\hbar})$ regardless of trajectory length.
Proliferation of diffractive orbits relative to geometric ones
is restricted to additional half-period contributions
by some orbits,in contrast to the billiard result ~\cite{sieber}, where any
fraction may contribute. We conclude that, in the accessible $\hbar$ range,
the diffractive effect remains small for the chaotic atom, but represents 
a substantial effect in the integrable case, on quantum timescales. 

In the KAM system, deviations 
qualitatively similar in form to the integrable case occur.
We do not attempt a rigorous analysis of systems with a mixed classical
phase space, but the numerics  support the notion that the
structure of the Berry-Tabor formula \cite{BT} is retained for near-integrable
KAM systems \cite{Toms}, and that our analysis could also find
application in such systems.

The preceding arguments are of course incomplete for the rigorous development
of a semiclassical limit, as we do not consider multiple scattering effects
such as the creation of the {\it combination orbits}. However,
we will see that in the numerically accessible regime (and indeed far
beyond experimental resolution) it is not necessary to consider such
effects to accurately reproduce the diffractive corrections to the spectral
statistics for $L\agt {\cal O}(1)$, up to a small constant correction.

We now analyze the case of a non-hydrogenic Stark atom. 
Since we will only be interested in long orbits, for which the uniformity
principle \cite{HOdA} can be invoked to connect period with action, we can
write the Berry-Tabor
formula as a sum over all periodic orbits $j$ (including retracing to
negative time) in the form:
\begin{equation}
{\cal N}^{\rm osc}_G (\kappa) =
-i\kappa^{3\over 2}{\Delta_\epsilon \over\Delta_\kappa}
\sum_j { A_j \over  S_j}
e^{i( S_j \kappa - \alpha_j \pi/2)}\ .
\label{Ng}
\end{equation}
Here $ S_j$, $\alpha_j$, are the scaled action and the Maslov index,
$ A_j$ is the scaled amplitude factor from \cite{BT},
$\Delta_\kappa^{-1} = \partial {\cal N}^s/\partial\kappa |_\epsilon$,
and $\Delta_\epsilon^{-1}=\partial{\cal N}^s/\partial\epsilon |_\kappa$.
Note that $\Delta_\kappa /\Delta_\epsilon$ scales as $\kappa$.

The 
contribution to the spectral staircase from the corresponding single-scatter
diffractive periodic orbits yields
a sum over the same set of periodic orbits as in Eq.(\ref{Ng}):
\begin{equation}
{\cal N}^{\rm osc}_D (\kappa) =
\sin \delta_0 \sqrt{\kappa} {\Delta_\epsilon \over\Delta_\kappa}
\sum_j C_j  A_j 
e^{i( S_j \kappa - \alpha_j \pi/2 \pm \delta_0)}
\label{Nd}
\end{equation}
where the phaseshift $\pm \delta_0$ is positive for forward time tracings,
and negative for negative time retracings. 

The rigidity formula (\ref{rig}), which is easily adapted for the problem in
hand where the {\it levels} are the $\kappa_i$, leads to integrals involving
the products $\langle {\cal N}_G {\cal N}_G \rangle_\kappa$,
$\langle {\cal N}_G {\cal N}_D \rangle_\kappa$ and 
$\langle {\cal N}_D {\cal N}_D \rangle_\kappa$. These products involve many
rapidly oscillating terms, which, following Berry \cite{Berry85}, we assume
to vanish upon averaging. This approximation is usually known as the
{\it diagonal approximation}, but here we retain different (off-diagonal)
trajectories with identical action, which occur due to diffraction.

The resulting equation for the rigidity can be written
\begin{equation}
\Delta^{\delta_0}(L,L_c)=L/15 + \Delta^{DG}(L,L_c) + \Delta^{DD} (L,L_c)\ ,
\label{rigd0}
\end{equation}
where
\begin{eqnarray}
\Delta^{DG}(L,L_c)=&-4\kappa^2\sin^2\delta_0
\left ({\Delta_\epsilon \over \Delta_\kappa}\right )^2 \sum^+_j C_j 
\int_0^{2\pi/\Delta_\kappa L_c} \cr
&\times d S
(A_j^2 / S)\delta( S- S_j)
G(\Delta_\kappa L  S/2)
\label{rigDG}
\end{eqnarray}
\begin{eqnarray}
\Delta^{DD}(L,L_c)=&2\kappa \sin^2\delta_0
\left ({\Delta_\epsilon \over \Delta_\kappa}\right )^2\sum^+_j C_j^2
\int_0^{2\pi/\Delta_\kappa L_c}\cr
&\times d S
A_j^2 \delta( S- S_j)
G(\Delta_\kappa L  S/2)\ ,
\label{rigDD}
\end{eqnarray}
the sum is now only over positive traversals,
and $G(y)=1-F^2(y)-3[F'(y)]^2$ ($F(y)=\sin y/y$), is Berry's
{\it orbit selection function}, which is similar to the step function
$\Theta(y-\pi)$. We have had to introduce an upper cut-off to the
integrals, which will be seen to diverge. This divergence
is a direct result of neglecting the rapidly oscillating terms, and will
be discussed further below. We now concentrate on evaluating the
integrands of (\ref{rigDG}) and (\ref{rigDD}). For long orbits, we invoke the
Hannay-Ozorio de Almeida sum rule \cite{HOdA}, which can be expressed as
$\lim_{ S \to\infty}
\sum_j A_j^2 \delta( S - S_j)=
(\Delta_\kappa/\Delta_\epsilon^2)/2\pi\kappa^3$
which is independent of $\kappa$.
We consider this to be accurate for $ S > S^*$, so that our result
for (\ref{rigd0})
will only be accurate for $L<L_{\rm max}=2\pi/\Delta_\kappa  S^*$.
Indeed $L_{\rm max}$ also marks the onset
of non-universal deviations of the non-diffractive result from the simple
$L/15$ dependence \cite{Berry85}. 
Inserting the sum rule into (\ref{rigDG}) and (\ref{rigDD}),
averaging over the distribution of $C_j$
and evaluating the integrals, leads to closed form
solutions that can be expressed in terms of special functions
\cite{Walker}. Here, we only write out the asymptotic approximations
to the solutions:
\begin{equation}
\Delta^{DG}(L,L_c)\sim -{2 \langle C_j \rangle
\sin^2\delta_0 \over \pi\kappa\Delta_\kappa}
[\ln {2\pi L\over L_c} + \gamma_E - {9\over 4}]
\label{asymrigDG}
\end{equation}
\begin{equation}
\Delta^{DD}(L,L_c)\sim {2 \langle C_j^2 \rangle
\sin^2\delta_0\over\kappa^2\Delta_\kappa^2} 
[L_c^{-1}-L^{-1}]
\ ,
\label{asymrigDD}
\end{equation}
where $\gamma_E$ is Euler's constant.$\kappa\Delta_\kappa$ is a constant,
independent of $\kappa$ and
$\sim -2 \epsilon$ for our Stark spectra.
The result is valid for $L>L_c$, and the ambiguity in $L_c$ leads to
an ambiguity in the constant term. For $L\ll L_c$ both $\Delta^{DD}$ and
$\Delta^{DG}$ vanish. To proceed, one should invoke the semiclassical
sum rule due to Berry \cite{Berry85}. It is natural to then identify
$L_c$ as $2\pi/\Delta_\kappa S_c$, ( eg $L_c \sim.94$ for the case in
Fig. ~\ref{Fig3}) where $S_c$ is the
point where the semiclassical and quantum asymptotes for the form factor
coincide. Since we have neglected both off-diagonal
corrections and higher order scattering contributions, we cannot expect to
be able to evaluate the constant term correctly.
In this case we can simply set $L_c$ to unity, and accept that our
result may not be accurate around $L\simeq 1$.
We note that in the GOE case it is not possible to evaluate the constant
term semiclassically either \cite{Berry85}.

The final result for the rigidity is then
\begin{equation}
\Delta^{\delta_0}(L)= \Delta^{0}(L)+{\sin^2\delta_0 \over  \epsilon}
 [{\langle C_j \rangle \over\pi} \ln L -
{\langle C_j^2\rangle \over 2 \epsilon L}  ] + \chi
\label{final}
\end{equation}
for $L_{\rm max}\agt L\agt 1$, where $\chi$ is the constant shift.
For $L\alt 1$, $\Delta^{\delta_0}(L) = L/15$.
The final formula hence
does not contain the effective Planck's constant,
proving the most surprising feature of Fig. ~\ref{Fig1}
namely that despite the $\hbar$-
dependent nature of the diffractive corrections, our quantal results
(below $L_{\rm max}$) are essentially $\hbar$-independent.

We find $\langle C_j \rangle \simeq 1.8$, and
$\langle C_j^2 \rangle \simeq  \langle C_j \rangle ^2$, 
for the $\epsilon=-3$ case.
We stress that these were not free parameters.
With $L_c \sim1.$ from (\ref{asymrigDG}) and (\ref{asymrigDD}) we 
find $\chi \sim .16$ for $\delta_0 =\pi /2$. However we estimate from the
quantal results that $\chi \sim 0.21$. The correction required can
be attributed to the neglect of both higher order scattering, and off-diagonal
contributions.

Our result  is compared with the fully quantal values shown in
Fig. ~\ref{Fig1}, and the quantal and semiclassical rigidities are
plotted in Fig. ~\ref{Fig3} .
The agreement (to within the small correction to $\chi$ ) is extremely good up
to $L_{\rm max}$, where the breakdown was expected.
The divergent nature of Eq.(\ref{final})
for $L<1$ is clearly seen. The $\sin^2\delta_0$ dependence has
been verified by considering several different defect values
(not shown) \cite{Walker}.

To summarise, we have combined the semiclassical theory of diffraction
and atomic Quantum Defect Theory
with the Berry-Tabor trace formula to give diffractive
corrections to the spectral rigidity of atoms in fields. We show that within a small 
constant shift, the semiclassical one-scatter results agree extremely accurately
with the fully quantal results.

The authors gratefully acknowledge helpful discussions 
with E Bogomolny, D Delande,
 D Ullmo, P. Braun and S. Owen,
and support by the EPSRC. We thank D Delande for pointing out the relation 
between $\kappa \Delta_{\kappa}$ and $\epsilon$.

\end{document}